\documentclass[useAMS,usenatbib]{mn2e}

\usepackage{caption}
\usepackage{graphicx}
\usepackage{textcomp}
\usepackage{latexsym}
\usepackage{varioref}
\usepackage{xspace}
\usepackage{makeidx}
\usepackage{verbatim}
\usepackage{tabularx}
\usepackage{epstopdf}
\usepackage{amsmath}
\usepackage{array}
\usepackage{units}
\usepackage{rotating}
\usepackage{float}

\def\Kepler{\textit{Kepler}}

\title[Broad-band timing properties of MV Lyrae]
{Broad-band timing properties of the accreting white dwarf MV Lyrae}
\author[S. Scaringi \textit{et al.}]
{S. Scaringi$^{1}$\thanks{E-mail: s.scaringi@astro.ru.nl}, E. K\"{o}rding$^{1}$, P. Uttley$^{2}$, P.J. Groot$^{1}$, C. Knigge$^{3}$, \newauthor M. Still$^{4,5}$, P. Jonker$^{6,1,7}$\\
$^{1}$Department of Astrophysics/IMAPP, Radboud University Nijmegen, P.O. Box 9010, 6500 GL Nijmegen, The Netherlands \\ 
$^{2}$Astronomical Institute ``Anton Pannekoek'', University of Amsterdam, Science Park 904, 1098XH, Amsterdam, The Netherlands \\
$^{3}$Department of Physics and Astronomy, University of Southampton, Highfield, Southampton, SO17 1BJ, UK \\ 
$^{4}$NASA Ames Research Center, Moffett Field, CA 94035, USA \\
$^{5}$Bay Area Environmental Research Institute, Inc., 560 Third St.West, Sonoma, CA 95476, USA \\
$^{6}$SRON, Netherlands Institute for Space Research, Sorbonnelaan 2, 3584 CA, Utrecht, The Netherlands \\
$^{7}$Harvard-Smithsonian Center for Astrophysics, 60 Garden Street, Cambridge, MA 02138, USA \\
}

\begin{document} 

\date{}

\pagerange{\pageref{firstpage}--\pageref{lastpage}} \pubyear{2012}

\maketitle

\label{firstpage}

\begin{abstract}
We present a broad-band timing analysis of the accreting white dwarf system MV Lyrae based on data obtained with the \Kepler\ satellite. The observations span 633 days at a cadence of 58.8 seconds and allow us to probe 4 orders of magnitude in temporal frequency. The modelling of the observed broad-band noise components is based on the superposition of multiple Lorentzian components, similar to the empirical modelling adopted for X-ray binary systems. We also present the detection of a frequency varying Lorentzian component in the lightcurve of MV Lyrae, where the Lorentzian characteristic frequency is inversely correlated with the mean source flux. Because in the literature similar broad-band noise components have been associated to either the viscous or dynamical timescale for different source types (accreting black holes or neutron stars), we here systematically explore both scenarios and place constraints on the accretion disk structure. In the viscous case we employ the fluctuating accretion disk model to infer parameters for the viscosity and disk scale height, and infer uncomfortably high parameters to be accommodated by the standard thin disk, whilst in the dynamical case we infer a large accretion disk truncation radius of $\approx10R_{WD}$. More importantly however, the phenomenological properties between the broad-band variability observed here and in X-ray binaries and Active Galactic Nuclei are very similar, potentially suggesting a common origin for the broad-band variability.   

\end{abstract}


\section{Introduction}

Compact interacting binaries (CBs) are close binary systems usually consisting of a late-type star that transfers material onto a black hole (BH), a neutron star (NS) or a white dwarf (WD) via Roche-lobe overflow. With an orbital period on the order of hours, the donor star transfers material through the L1 point, which forms an accretion disk surrounding the compact object. The dynamics and physics governing the flow of matter accreting onto the compact objects is, however, still debated. The accretion disks in BH and NS binary systems (X-ray binaries or XRBs) emit most of their radiation in X-rays, whilst the accretion disks in accreting white dwarfs (cataclysmic variables or CVs) emit mostly in the optical/UV wavebands. This is a consequence of the gravitational potential well created by the central compact object: for CVs the inner-most edge of the accretion disk sits at a few thousand gravitational radii, whilst for galactic BH and NS the inner disk reaches down to a few gravitational radii.

Most CVs and XRBs are highly variable sources in X-rays and/or optical/UV. This variability has been associated with the accretion disk from the characteristic frequencies observed in the power spectral density (PSD). For example, the presence of periodic modulations slightly longer than the orbital period can be associated with positive superhumps caused by a tidal deformation in the accretion disk (\citealt{whitehurst,osaki,lubow}). This has been modelled and observed in CVs (\citealt{montgomery,wood}), as well as in XRBs (\citealt{dono}).  

Another example of similar timing characteristics between CVs and XRBs is the presence of both dwarf nova oscillations (DNOs) and quasi-periodic oscillations (QPOs). In both cases DNOs/QPOs appear in the PSD at a few tens of mHz (\citealt{warner_qpo,pretorius_qpo}), whilst for XRBs they appear at few hectoHz, and are referred to as lower kHz oscillations (\citealt{belloni}). The phenomenological similarity between the QPOs and DNOs observed in CVs and XRBs was first noted by \cite{mauche} and \cite{warner_qpo}, where the ratio of periods is $P_{QPO}/P_{DNO} \approx 15$, and holds over 6 orders of magnitude in temporal frequency. The physical reason for this relation is not fully understood, although some suggestions have been proposed (\citealt{belloni}), all involving accretion disk dynamics, but none seem to be able to explain all observations consistently (\citealt{warner_qpo,pretorius_qpo}).

In addition to periodic or quasi-periodic signals, both CVs and XRBs possess an intrinsic aperiodic broad-band noise continuum, generally described as flickering in CVs. This red noise component produces a continuum that rises towards low frequencies. In the XRB context, this is usually modelled a by simple power-law (e.g. $1/f$) or by a sum of broad Lorentzian components (\citealt{belloni}), with a characteristic break at $\approx 10^{-3}$ Hz for CVs and $\approx 1$ Hz for XRBs (\citealt{revn1,belloni}). This component has been shown to display a linear rms-flux relation in XRBs and Active Galactic Nuclei (AGN, \citealt{uttley1}), where the root-mean-square (rms) variability linearly scales with flux over a wide range of timescales. More recently, the same relation has been also observed in the CV system MV Lyrae, albeit at lower frequencies (\citealt{scaringi}, hereafter Paper~I). The detection of the rms-flux relation in both CVs and XRBs strongly suggests that the broad-band variability originates within the accretion disk. Under certain assumptions about the origin of the variability and the physics of the mass-transfer process, the characteristics of the broad-band noise can be used to constrain the strength of the effective viscosity that drives the mass transfer through the disk (\citealt{ID1,ID2}). 

Motivated by the similarities between the periodic and aperiodic signals in CVs and XRBs, we here study the broad-band temporal frequency spectrum of the CV MV Lyrae in search of further similarities between the two classes of objects. MV Lyrae is one of 14 known CVs in the \Kepler\ field-of-view (FOV), and is classified as being a VY Scl novalike system, spending most of its time in a high state ($V\approx12-13$), but occasionally (every few years) undergoing short-duration (weeks to months) drops in brightness ($V\approx16-18$, \citealt{hoard}). The reason for these sudden drops in luminosity is not clear, but \cite{LP} have suggested star spots from the donor covering the L1 point inhibiting mass transfer. It is known however that MV Lyrae has an extremely low mass transfer rate at its minimum brightness ($3\times10^{-13}M_{\odot}/yr$, \citealt{linnell,hoard}), where the WD is detected at $V\approx18$ and dominates the emitted light. Furthermore, the orbital period of $3.19$ hours has been determined for the system, as well as a low inclination of $i\sim11^o-13^o$ (\citealt{SPT95}). 

In Paper~I we analysed the high frequency (tens of minutes) broad-band variability for MV Lyrae with data obtained with the \Kepler\ satellite. In this paper we will analyse the same lightcurve at lower temporal frequencies, with particular emphasis on the varying broad-band components during the observation. MV Lyrae has been observed with \Kepler\ during a low-to-high luminosity transition, when its optical emission originates mostly from its nearly face-on disk. 

In section \ref{sec:data}, we briefly describe the \Kepler\ data acquisition, and the procedure we use to construct broad-band power spectral densities (PSDs). We fit the PSDs with a combination of Lorentzian shaped functions to characterise the observed broad-band noise components. Section \ref{sec:results} shows our results, with particular emphasis on one frequency-varying broad-band noise component, which appears to correlate with the mean source flux. Finally, in section \ref{sec:discussion}, we consider whether the inferred broad-band noise components can be associated with fluctuations on either the dynamical or viscous time-scales in the accretion disk. As we shall see, both of these interpretations lead to uncomfortable implications for the accretion disk structure.

\section{Data analysis}\label{sec:data}

\begin{figure*}
\includegraphics[width=\textwidth]{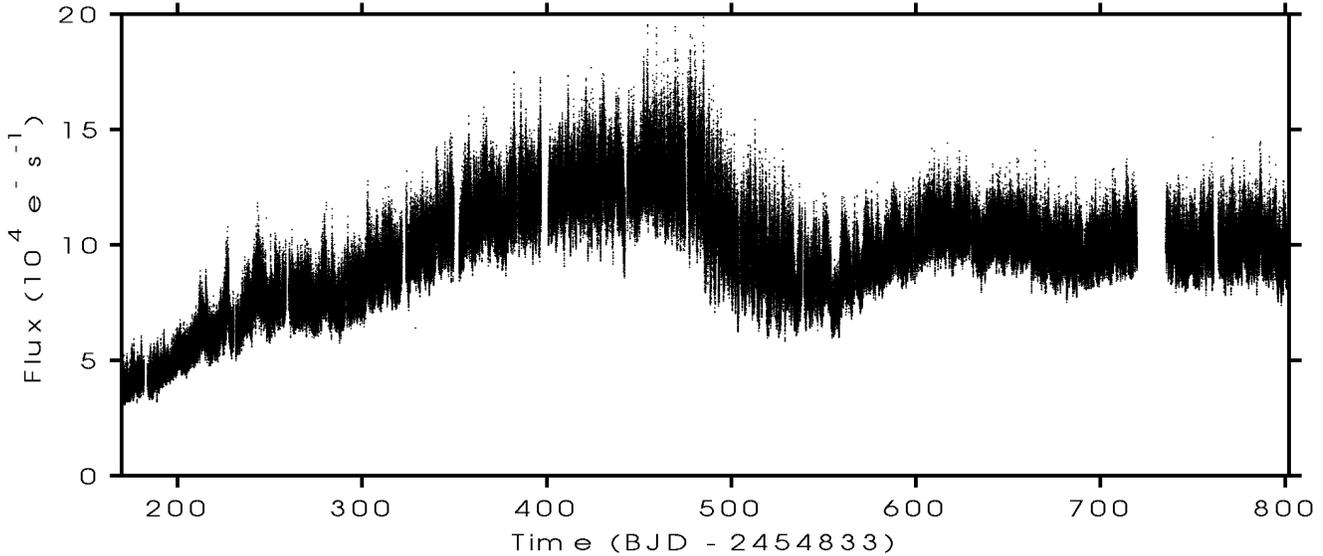}	
\caption{MV Lyrae light curve obtained by the \Kepler\ satellite in short cadence mode (58.8 seconds). }
\label{fig:1}
\end{figure*}

The MV Lyrae lightcurve\footnote{The \Kepler\ response function for the photometry covers the wavelength range $4000-9000$\AA.} was provided to us by the Science Operations Centre in reduced and calibrated form after being run through the standard data reduction pipeline (\citealt{jenkins}). Similarly to Paper~I, we only consider the Single Aperture Photometry (SAP) lightcurve. Data gaps occasionally occur due to \Kepler\ entering anomalous safe modes (see the \Kepler\ Data Characteristics Handbook\footnote{http://archive.stsci.edu/kepler/documents.html}). Here, we make no attempt to correct these artifacts, but simply remove them from the light curve. Fig. \ref{fig:1} shows the short cadence (58.8 seconds; \citealt{gilliland}), barycentre corrected, lightcurve for MV Lyrae obtained during the first eight quarters of \Kepler\ operations. The lightcurve spans an interval of 633 days. The visible data gaps present in the lightcurve are due to the artifacts described in Paper~I as well as the monthly data down-links.

\begin{figure}
\includegraphics[width=0.45\textwidth]{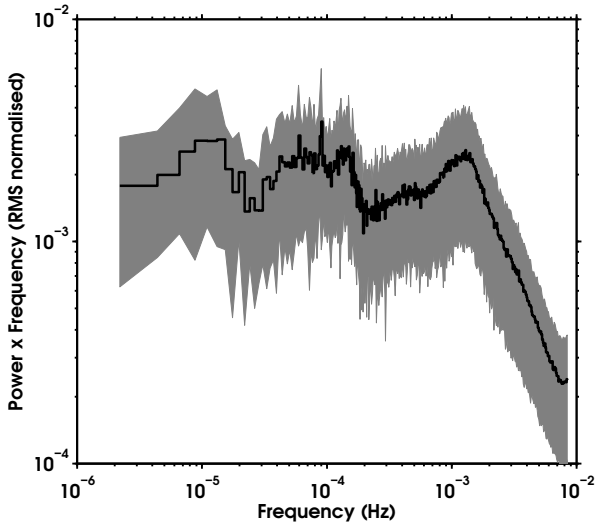}	
\caption{Time-averaged PSD of MV Lyrae. The shaded region shows the 1-$\sigma$ width of the distribution spanned by the 118 single 5.275 day PSDs, whilst the thick line shows their average.}
\label{fig:2}
\end{figure}

In order to obtain the time-averaged PSDs, we have split the lightcurve in Fig. \ref{fig:1} into 120 non-overlapping segments, each covering 5.275 days. This timescale has been chosen to probe the lowest frequencies without being affected by low frequency power generated by the long term trends in flux during the observation (see Fig. \ref{fig:1}). We computed the Fast-Fourier transform (FFT) of 118 segments, ignoring 2 segment due to the large data gap at $\approx750$ days in Fig. \ref{fig:1}. We further applied the rms normalisation of \cite{miyamoto} so that the square root of the integrated PSD power over a specific frequency range yields the root-mean-square (rms) variability. After computing the normalised PSDs for each segment we obtain the intrinsic spread in each frequency bin by computing the standard deviation of the 118 PSDs, shown in Fig. \ref{fig:2}. 

\begin{figure}
\includegraphics[width=0.5\textwidth]{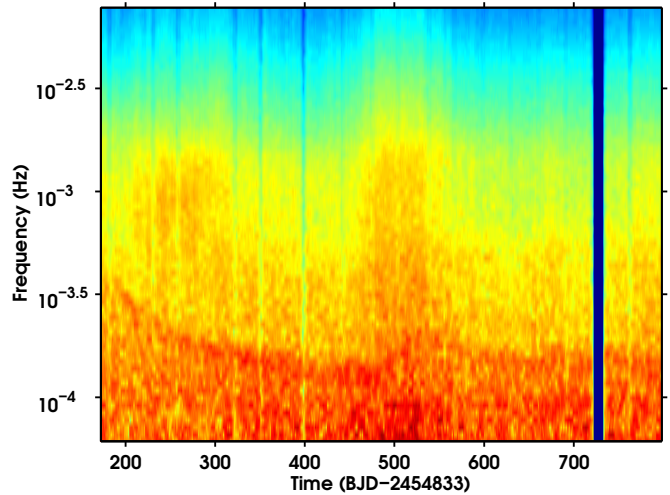}
\caption{Rms-normalised dynamic PSD of the MV Lyrae lightcurve shown in Fig. \ref{fig:1}. Each PSD has been computed on a 5.275 days segment, with a $50\%$ overlap.}
\label{fig:3}
\end{figure}

At first glance we note large scatter in the time-averaged PSD, caused by the varying PSD shape and normalisation during the 633 day observation. Furthermore many incoherent features are clear from Fig. \ref{fig:2}. Specifically, we see a clear broad Lorentzian-shaped feature peaking at about $10^{-3}$ Hz, briefly discussed in Paper~I, but also excess power at about $10^{-4}$ and $10^{-5}$ Hz. It is hard to determine whether the excess power is caused by a single intrinsically broad feature or a blend of many. In order to shed light on which of these interpretations is correct we have produced a dynamic FFT of the lightcurve. This is shown in Fig. \ref{fig:3}, where we have again computed rms-normalised PSDs on 5.275 day segments, but this time with a $50\%$ overlap, for clarity. The apparent gaps are caused by data gaps in the lightcurve shown in Fig. \ref{fig:1}. Fig. \ref{fig:3} suggest that the excess power between $10^{-3.5}$Hz and $10^{-4}$Hz is produced by a single, frequency-varying QPO whose peak frequency is initially near $10^{-3.5}$Hz and then slowly declines on a time-scale of $\approx250$ days. A comparison with the lightcurve in Fig. \ref{fig:1} also suggests a possible correlation between the peak frequency of this QPO and the mean flux of the system. We will return to this correlation in section \ref{sec:results}.

\subsection{PSD fitting}
In order to quantify and fit the frequency-varying QPO(s), we first produced 118 rms-normalised PSDs, each from 5.275-day, non-overlapping data segments. Next, we averaged 5 consecutive PSDs (except in one case where we averaged 3 due to a large data gap in the lightcurve) and binned these into $\approx550$ equally spaced frequency intervals. The bin size is somewhat arbitrary, but has been chosen as a compromise to increase the signal-to-noise in each bin while retaining a high frequency resolution. We further obtained the intrinsic scatter in each frequency bin by calculating the standard deviation of the 5 PSDs used. This intrinsic scatter is larger than the statistical errors, and we thus adopt the intrinsic scatter as a conservative estimate of the errors of each frequency bin, which also takes into account fluctuations in the intrinsic PSD shape. This procedure resulted in 24 time-averaged PSDs.

The fit to each PSD was carried out by starting with a simple broken power-law fit, to which we added successive Lorentzian component until the reduced $\chi^2$ was near unity. The shape of the Lorentzians was taken from \cite{belloni}, 

\begin{equation}
P(\nu) = {{r^2}\Delta \over \pi} {1 \over { {\Delta^2} + {(\nu - \nu_0)^2} } },
\label{equ:1}
\end{equation}

\noindent where $r$ is the integrated fractional rms and $\Delta$ is the half-width at half-maximum. With this definition the Lorentzians attain their maximum power in $\nu P(\nu)$ at the characteristic frequency

\begin{equation}
\nu_{max} = {1 \over t_{max}} =\sqrt{{\nu_0^2} + {\Delta^2}}.
\label{equ:2}
\end{equation}

\noindent The broken power-law takes the form
\begin{equation}
P(\nu) = {A\nu^{-1} \over {1+ {\nu \over \nu_{br}}} },
\label{equ:3}
\end{equation}

\noindent where $A$ is the normalisation and $\nu_{br}$ the break frequency. With this definition the PSD shape has a slope of $-1$ up to $\nu_{br}$, where the slope then changes to $-2$, similarly to the PSDs observed in some AGN and XRBs (\citealt{summons}). In our analysis we are mainly interested in the properties of the fitted Lorentzian, and use the low-frequency power-law model to account for any systematic excess power generated by the long term trends (such as the steady flux increase) observed in the lightcurve shown in Fig. \ref{fig:1}.

\begin{figure*}
\includegraphics[width=0.48\textwidth]{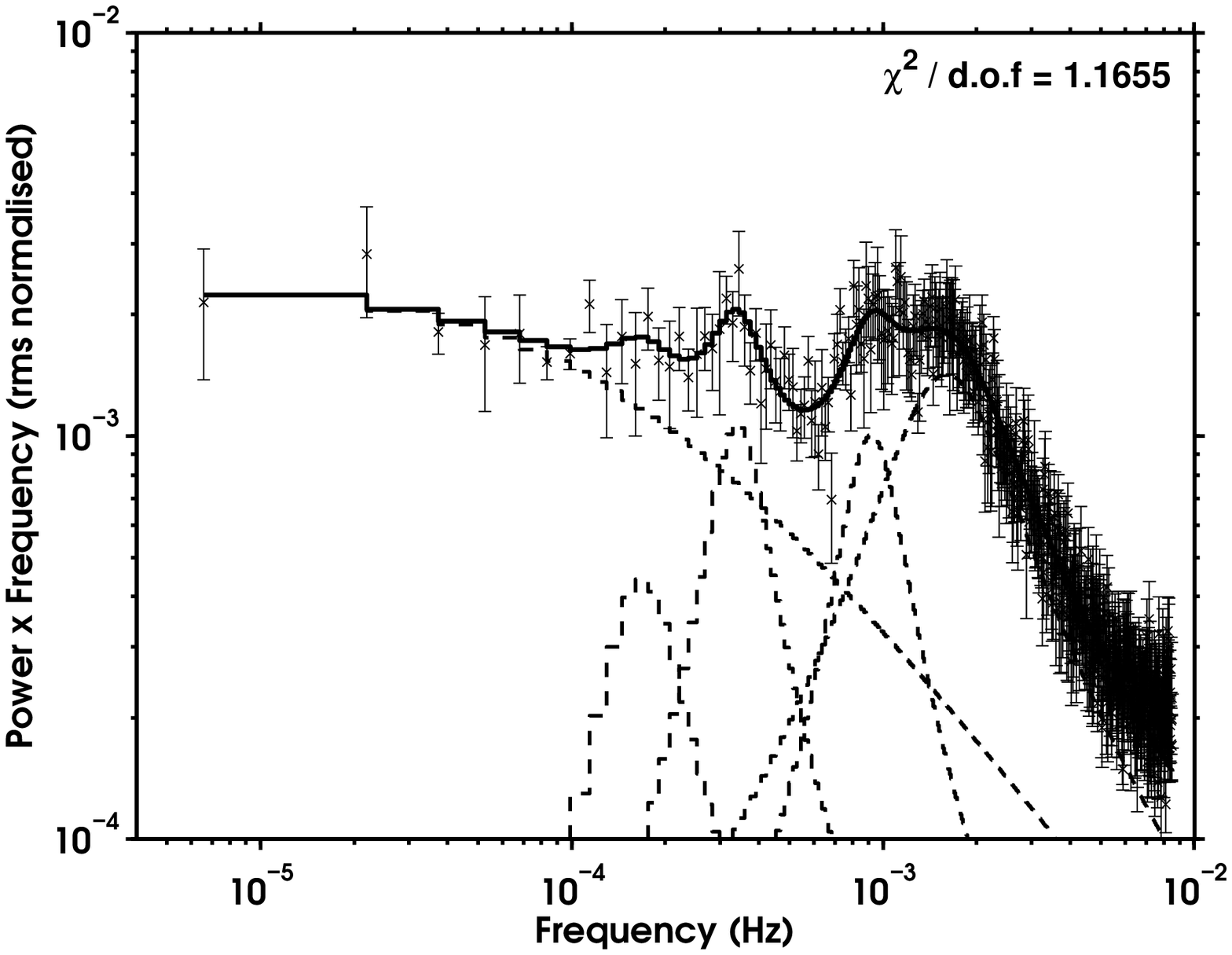} \includegraphics[width=0.48\textwidth]{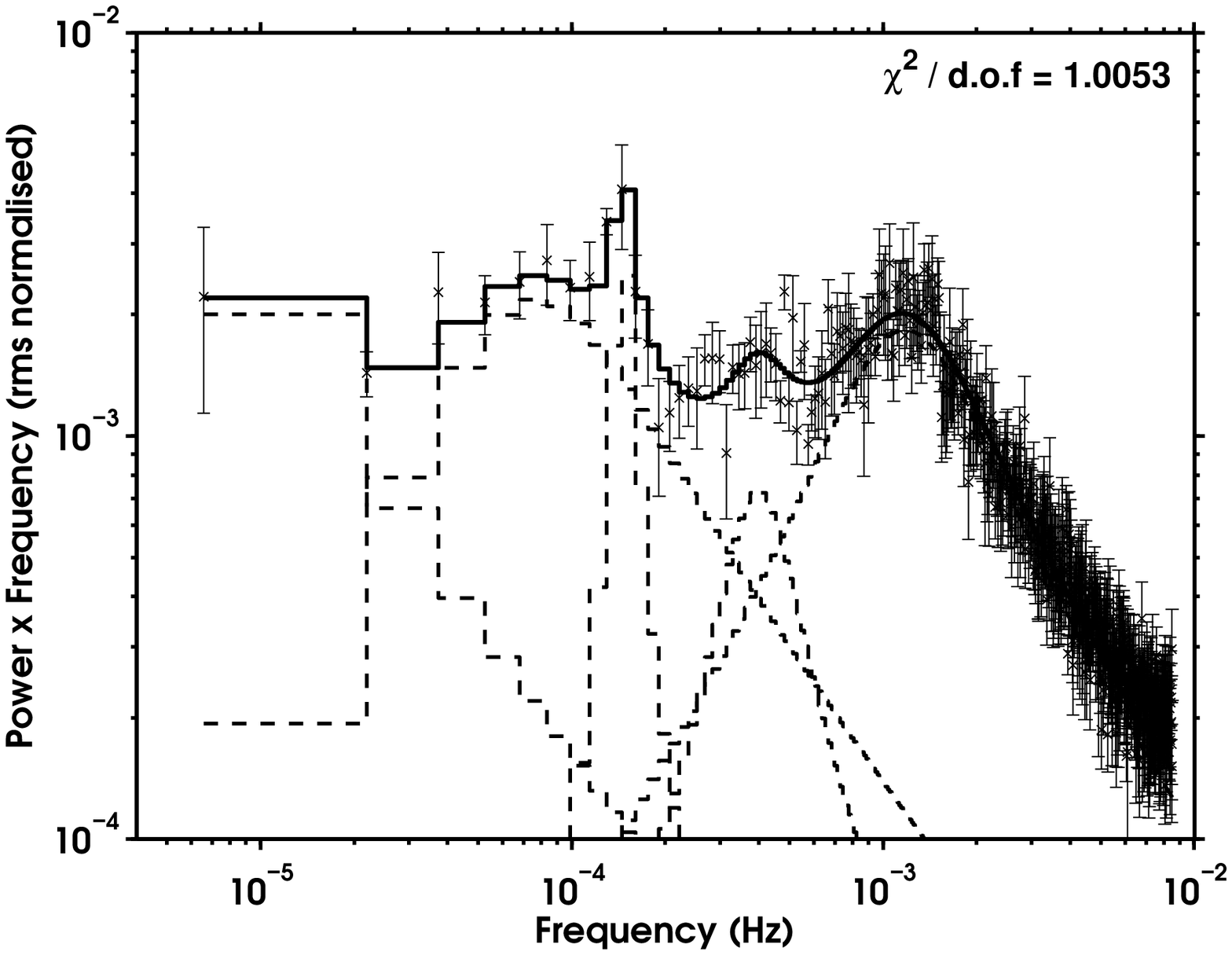}
\includegraphics[width=0.48\textwidth]{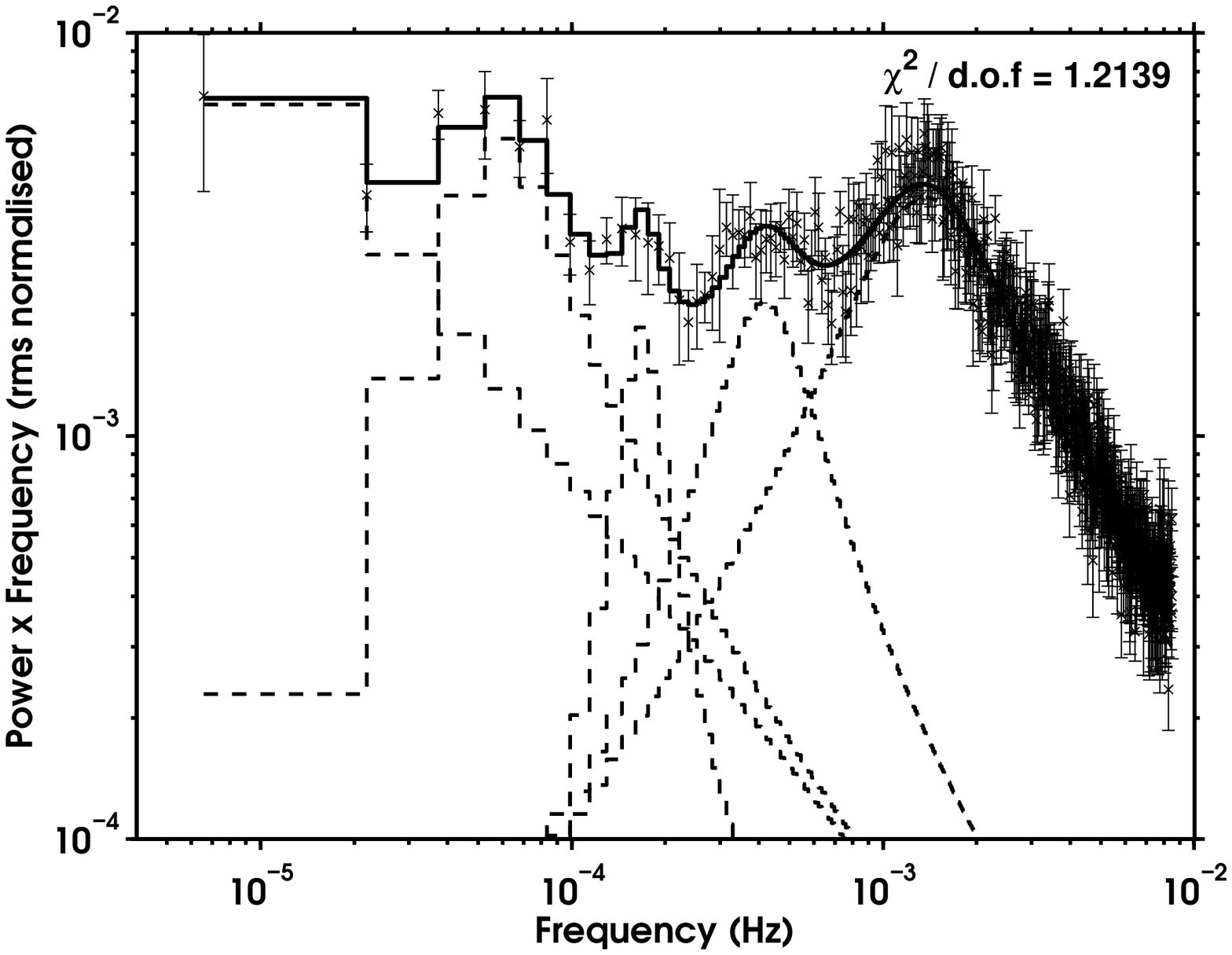} \includegraphics[width=0.48\textwidth]{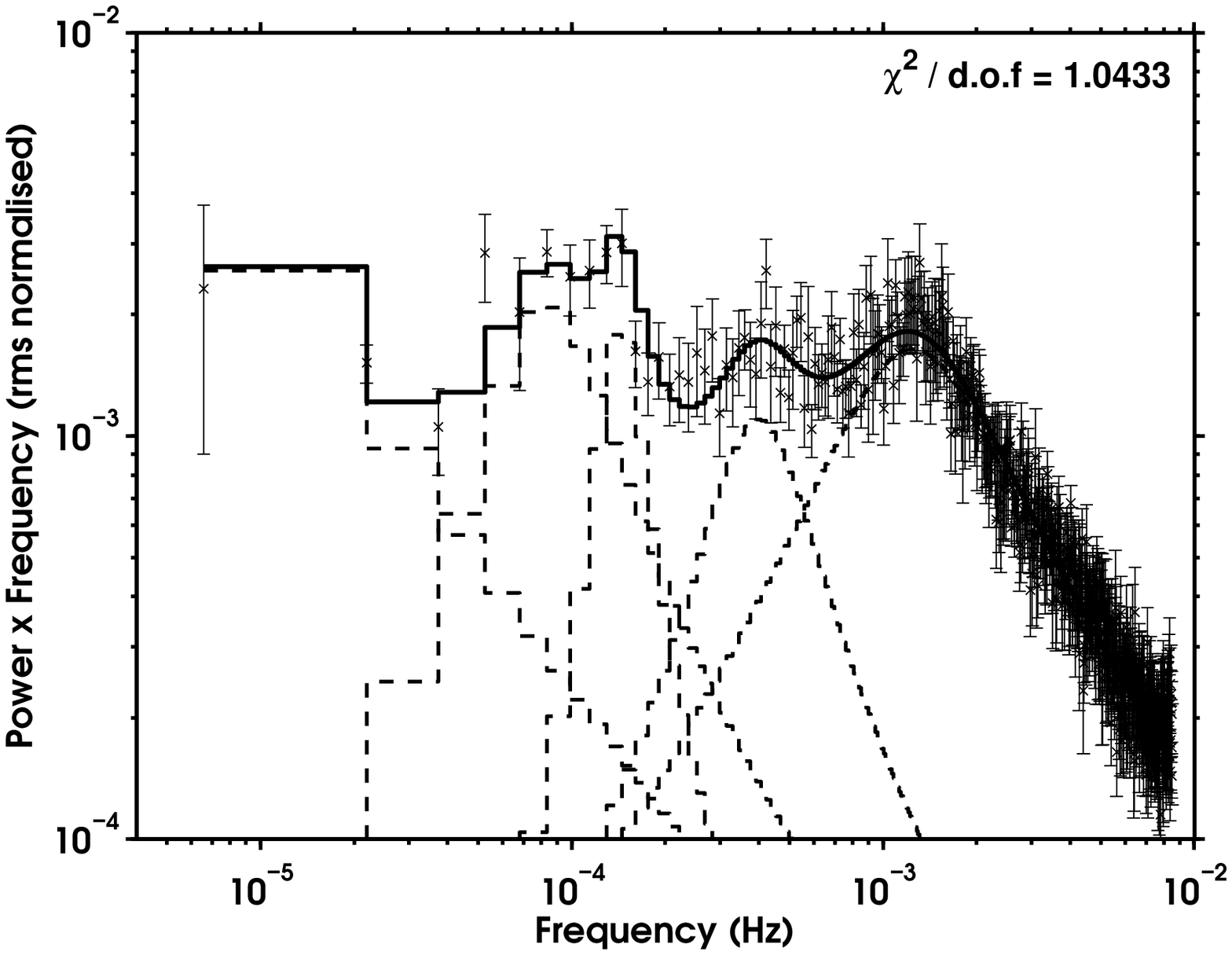}
\caption{PSDs in $\nu P_{\nu}$ form for MV Lyrae, also showing the best-fit model components. The PSDs have been created by averaging 5 5.275 day PSDs. From top-left to bottom-right the central observation times are respectively $182.7$, $367.3$, $525.5$, $789.1$ in BJD-$2454833$.}
\label{fig:4}
\end{figure*}

\begin{table*}\small
\centering
\begin{tabular}{l l l l l l l l}
\hline
\hline
            &                     &           &    $L_1$      &             &           &    $L_2$      &             \\
\hline  
Time        & Mean flux           &  $r$      &  $\nu_{0}$    & $\Delta$    & $r$       &  $\nu_{0}$    & $\Delta$    \\
BJD-2554833 & $10^{4}e^{-}s^{-1}$ & $10^{-2}$ &  $10^{-4}Hz$  & $10^{-4}$Hz & $10^{-2}$ &  $10^{-4}$Hz  & $10^{-4}$Hz  \\
\hline
182.69  &  4.29  &  4.77$\pm$0.49   &   13.97$\pm$1.44   &   7.62$\pm$0.61   &   2.79$\pm$0.71   &   8.83$\pm$0.36   &   2.20$\pm$0.88   \\   
209.06  &  5.69  &  5.70$\pm$1.47   &   11.21$\pm$2.43   &   8.06$\pm$1.00   &   4.11$\pm$2.06   &   7.53$\pm$0.54   &   3.27$\pm$1.75   \\   
235.43  &  7.20  &  8.67$\pm$0.39   &   8.23$\pm$0.49   &   5.87$\pm$0.23   &   1.00$\pm$1.17   &   6.96$\pm$0.60   &   0.69$\pm$1.38   \\   
261.79  &  7.70  &  8.48$\pm$0.19   &   8.80$\pm$0.25   &   5.09$\pm$0.21   &   0.94$\pm$0.32   &   4.80$\pm$0.11   &   0.18$\pm$0.15   \\   
288.16  &  7.85  &  8.54$\pm$0.21   &   9.68$\pm$0.33   &   5.96$\pm$0.20   &   2.64$\pm$0.57   &   4.24$\pm$0.26   &   1.44$\pm$0.58   \\   
314.52  &  9.10  &  7.48$\pm$0.23   &   9.01$\pm$0.59   &   8.09$\pm$0.30   &   2.80$\pm$0.46   &   4.09$\pm$0.17   &   1.17$\pm$0.35   \\   
340.88  &  10.32  &  6.23$\pm$0.34   &   8.43$\pm$0.89   &   7.61$\pm$0.39   &   2.83$\pm$0.87   &   3.66$\pm$0.39   &   1.54$\pm$0.71   \\   
367.25  &  11.41  &  6.20$\pm$0.19   &   9.36$\pm$0.52   &   7.04$\pm$0.31   &   2.65$\pm$0.42   &   3.73$\pm$0.19   &   1.16$\pm$0.33   \\   
393.62  &  12.20  &  3.94$\pm$0.27   &   10.23$\pm$0.36   &   4.58$\pm$0.52   &   0.99$\pm$0.14   &   3.58$\pm$0.04   &   0.10$\pm$0.04   \\   
419.98  &  12.71  &  5.76$\pm$0.14   &   9.00$\pm$0.42   &   6.99$\pm$0.28   &   2.63$\pm$0.31   &   3.45$\pm$0.11   &   0.86$\pm$0.19   \\   
446.35  &  12.90  &  6.35$\pm$0.22   &   8.52$\pm$0.69   &   8.02$\pm$0.39   &   2.57$\pm$0.39   &   3.24$\pm$0.16   &   1.00$\pm$0.30   \\   
472.72  &  12.92  &  9.12$\pm$0.31   &   8.03$\pm$0.70   &   8.59$\pm$0.32   &   3.12$\pm$0.66   &   3.68$\pm$0.22   &   1.21$\pm$0.44   \\   
499.09  &  10.40  &  9.69$\pm$0.45   &   10.46$\pm$0.67   &   8.32$\pm$0.31   &   5.75$\pm$1.15   &   2.81$\pm$0.52   &   2.36$\pm$0.77   \\   
525.45  &  9.05  &  8.80$\pm$0.28   &   11.17$\pm$0.57   &   7.83$\pm$0.37   &   5.20$\pm$0.49   &   3.76$\pm$0.17   &   1.56$\pm$0.30   \\   
551.82  &  8.41  &  7.32$\pm$0.22   &   10.94$\pm$0.67   &   8.50$\pm$0.48   &   2.40$\pm$0.39   &   4.33$\pm$0.15   &   0.75$\pm$0.27   \\   
578.19  &  9.40  &  5.39$\pm$0.23   &   11.13$\pm$0.62   &   7.61$\pm$0.34   &   4.29$\pm$0.39   &   3.57$\pm$0.20   &   2.00$\pm$0.34   \\   
604.55  &  10.48  &  4.77$\pm$0.37   &   11.38$\pm$1.18   &   8.49$\pm$0.56   &   3.96$\pm$0.52   &   4.22$\pm$0.27   &   2.41$\pm$0.57   \\   
630.92  &  10.53  &  5.12$\pm$0.28   &   9.71$\pm$0.94   &   8.96$\pm$0.40   &   2.88$\pm$0.64   &   3.47$\pm$0.41   &   2.10$\pm$0.73   \\   
657.28  &  10.35  &  4.84$\pm$0.20   &   10.81$\pm$0.83   &   8.33$\pm$0.53   &   2.74$\pm$0.41   &   4.13$\pm$0.18   &   1.19$\pm$0.32   \\   
683.65  &  9.71  &  5.12$\pm$0.32   &   10.70$\pm$1.11   &   8.74$\pm$0.52   &   3.55$\pm$0.60   &   4.20$\pm$0.29   &   1.93$\pm$0.52   \\   
710.01  &  10.18  &  4.82$\pm$0.46   &   8.94$\pm$1.72   &   9.43$\pm$0.65   &   2.61$\pm$0.74   &   4.31$\pm$0.39   &   2.11$\pm$0.89   \\   
736.38  &  9.98  &  5.19$\pm$0.30   &   9.49$\pm$1.15   &   8.94$\pm$0.51   &   2.64$\pm$0.49   &   4.28$\pm$0.25   &   1.56$\pm$0.52   \\   
762.74  &  10.06  &  5.71$\pm$0.29   &   8.86$\pm$1.00   &   8.45$\pm$0.45   &   2.53$\pm$0.48   &   4.45$\pm$0.19   &   1.20$\pm$0.41   \\   
789.11  &  10.02  &  6.10$\pm$0.35   &   9.53$\pm$1.01   &   7.92$\pm$0.50   &   3.80$\pm$0.58   &   3.58$\pm$0.24   &   1.56$\pm$0.45   \\   

\end{tabular}
\\

\begin{tabular}{l l l l l l l l l}
\\
\hline
\hline
\hline
          &  $L_3$       &             &            &  $L_4$        &              &            &             &                 \\
\hline  
$r$       &  $\nu_{0}$   & $\Delta$    & $r$        &  $\nu_{0}$    & $\Delta$     &  $A$       & $\nu_{br} $ & $\chi^2$(d.o.f) \\
$10^{-2}$ &  $10^{-4}$Hz & $10^{-5}$Hz & $10^{-2}$  &  $10^{-4}$Hz  & $10^{-5}$Hz  &  $10^{-3}$ & $10^{-5}$Hz &                 \\
\hline
2.81$\pm$0.37   &   3.23$\pm$0.14   &   7.71$\pm$2.03   &   2.11$\pm$0.87   &   1.53$\pm$0.17   &   5.06$\pm$3.63   &   2.32$\pm$0.07   &   16.22$\pm$4.46   &   628.22(539)   \\   
2.33$\pm$0.46   &   2.64$\pm$0.04   &   2.49$\pm$0.88   &   4.60$\pm$0.86   &   1.42$\pm$0.15   &   10.11$\pm$3.30   &   2.15$\pm$0.11   &   3.41$\pm$0.76   &   572.12(539)   \\   
2.04$\pm$0.62   &   1.94$\pm$0.06   &   2.75$\pm$1.34   &   6.58$\pm$1.32   &   0.48$\pm$0.24   &   8.55$\pm$2.32   &   10.64$\pm$3.24   &   0.35$\pm$0.17   &   634.09(539)   \\   
2.35$\pm$0.27   &   1.81$\pm$0.02   &   1.41$\pm$0.45   &   7.60$\pm$0.79   &   0.22$\pm$0.17   &   8.33$\pm$1.13   &   5.00             &   0.27$\pm$0.02   &   573.31(540)   \\   
2.88$\pm$0.16   &   1.63$\pm$0.02   &   1.53$\pm$0.21   &   2.04$\pm$0.28   &   1.08$\pm$0.02   &   1.13$\pm$0.60   &   1.85$\pm$0.08   &   7.51$\pm$1.54   &   678.57(539)   \\   
2.17$\pm$0.19   &   1.58$\pm$0.02   &   0.85$\pm$0.20   &   9.02$\pm$0.61   &   0.17$\pm$0.08   &   5.55$\pm$0.44   &   9.89$\pm$0.51   &   0.10            &   491.48(540)   \\   
2.95$\pm$0.20   &   1.47$\pm$0.01   &   0.87$\pm$0.14   &   8.89$\pm$2.06   &   0.24$\pm$0.16   &   4.60$\pm$0.47   &   5.39$\pm$37.38   &   0.18$\pm$1.59   &   620.69(539)   \\   
2.78$\pm$0.22   &   1.39$\pm$0.01   &   1.10$\pm$0.31   &   8.41$\pm$0.38   &   0.40$\pm$0.03   &   5.67$\pm$0.54   &   15.15$\pm$0.48   &   0.10            &   542.88(540)   \\   
2.35$\pm$0.13   &   1.27$\pm$0.02   &   0.99$\pm$0.14   &   2.76$\pm$0.19   &   0.59$\pm$0.01   &   1.35$\pm$0.31   &   1.47$\pm$0.03   &   68.14$\pm$11.31   &   646.78(539)   \\   
1.68$\pm$1.90   &   1.26$\pm$0.05   &   0.29$\pm$0.99   &   7.48$\pm$0.76   &   0.40$\pm$0.08   &   5.52$\pm$0.52   &   7.02$\pm$2.73   &   0.46$\pm$0.31   &   652.78(539)   \\   
3.00$\pm$0.27   &   1.24$\pm$0.01   &   1.14$\pm$0.30   &   6.34$\pm$0.37   &   0.57$\pm$0.02   &   3.45$\pm$0.44   &   3.80            &   1.18$\pm$0.09   &   571.55(540)   \\   
3.84$\pm$3.33   &   1.22$\pm$0.01   &   0.34$\pm$0.89   &   8.44$\pm$1.31   &   0.43$\pm$0.07   &   4.07$\pm$0.66   &   5.01$\pm$1.35   &   1.14$\pm$0.92   &   611.36(539)   \\   
2.48$\pm$0.47   &   1.46$\pm$0.04   &   1.88$\pm$0.63   &   7.68$\pm$0.36   &   0.55$\pm$0.01   &   1.87$\pm$0.15   &   11.23$\pm$1.64   &   0.72$\pm$0.21   &   503.37(539)   \\   
3.17$\pm$0.41   &   1.58$\pm$0.04   &   2.73$\pm$0.77   &   9.50$\pm$0.59   &   0.45$\pm$0.02   &   2.51$\pm$0.25   &   15.80$\pm$4.22   &   0.48$\pm$0.22   &   654.27(539)   \\   
1.99$\pm$0.63   &   1.87$\pm$0.08   &   2.31$\pm$1.45   &   7.78$\pm$0.93   &   0.60$\pm$0.08   &   5.35$\pm$0.96   &   7.09$\pm$1.44   &   0.82$\pm$0.37   &   663.23(539)   \\   
2.41$\pm$0.24   &   1.60$\pm$0.03   &   2.34$\pm$0.41   &   4.62$\pm$0.27   &   0.64$\pm$0.02   &   2.98$\pm$0.35   &   3.41$\pm$0.40   &   0.90$\pm$0.23   &   561.41(539)   \\   
3.27$\pm$0.53   &   1.34$\pm$0.04   &   2.85$\pm$0.63   &   5.70$\pm$0.89   &   0.54$\pm$0.05   &   3.75$\pm$0.95   &   1.46$\pm$1.24   &   0.57$\pm$0.94   &   657.11(539)   \\   
1.92$\pm$0.23   &   1.42$\pm$0.02   &   1.21$\pm$0.30   &   6.68$\pm$0.32   &   0.55$\pm$0.03   &   5.31$\pm$0.50   &   1.90            &   0.47$\pm$0.03   &   459.03(540)   \\   
2.40$\pm$10.08   &   1.39$\pm$0.02   &   0.21$\pm$2.04   &   9.51$\pm$1.15   &   0.24$\pm$0.12   &   6.18$\pm$0.47   &   1.02$\pm$6.46   &   0.32$\pm$3.10   &   589.60(539)   \\   
1.90$\pm$0.59   &   1.32$\pm$0.04   &   1.94$\pm$0.94   &   7.52$\pm$0.81   &   0.48$\pm$0.06   &   5.35$\pm$0.80   &   1.50$\pm$0.66   &   0.96$\pm$1.12   &   556.03(539)   \\   
2.23$\pm$3.48   &   1.38$\pm$0.01   &   0.30$\pm$1.34   &   5.51$\pm$0.57   &   0.75$\pm$0.03   &   3.32$\pm$0.56   &   1.06$\pm$0.09   &   5.48$\pm$4.20   &   565.27(539)   \\   
3.20$\pm$13.70   &   1.54$\pm$0.01   &   0.15$\pm$1.38   &   8.13$\pm$0.89   &   0.36$\pm$0.07   &   4.22$\pm$0.36   &   5.01$\pm$16.93   &   0.19$\pm$0.87   &   602.86(539)   \\   
2.21$\pm$0.34   &   1.41$\pm$0.03   &   1.74$\pm$0.55   &   8.29$\pm$1.13   &   0.32$\pm$0.11   &   5.30$\pm$0.54   &   5.01$\pm$17.83   &   0.19$\pm$0.89   &   538.07(539)   \\   
3.21$\pm$0.63   &   1.34$\pm$0.04   &   2.29$\pm$0.70   &   5.77$\pm$0.72   &   0.68$\pm$0.03   &   3.55$\pm$0.90   &   10.56$\pm$6.00   &   0.21$\pm$0.16   &   562.37(539)   \\   

\end{tabular}

\caption{Best-fit parameters for the 24 time-averaged PSDs of MV Lyrae. In cases where the fit parameters were unconstrained no error bars are given.} \label{tab:1} 
\end{table*} 

\section{Results}\label{sec:results}
We fit each of the 24 PSDs individually, taking into account the Poisson noise level. We find that all PSDs require 4 Lorentzians. Of course, this does not exclude the possibility that more components could be present in the data, hidden beneath the noise level. Fig. \ref{fig:4} shows some examples of our fits, whilst Table \ref{tab:1} shows the resulting best-fit parameters for all our fits, where we labelled our Lorentzians from 1 to 4 in ascending order of frequency. In a few cases when two Lorentzians were highly blended the fits resulted in some parameters being unconstrained.

\begin{figure*}
\includegraphics[width=0.48\textwidth]{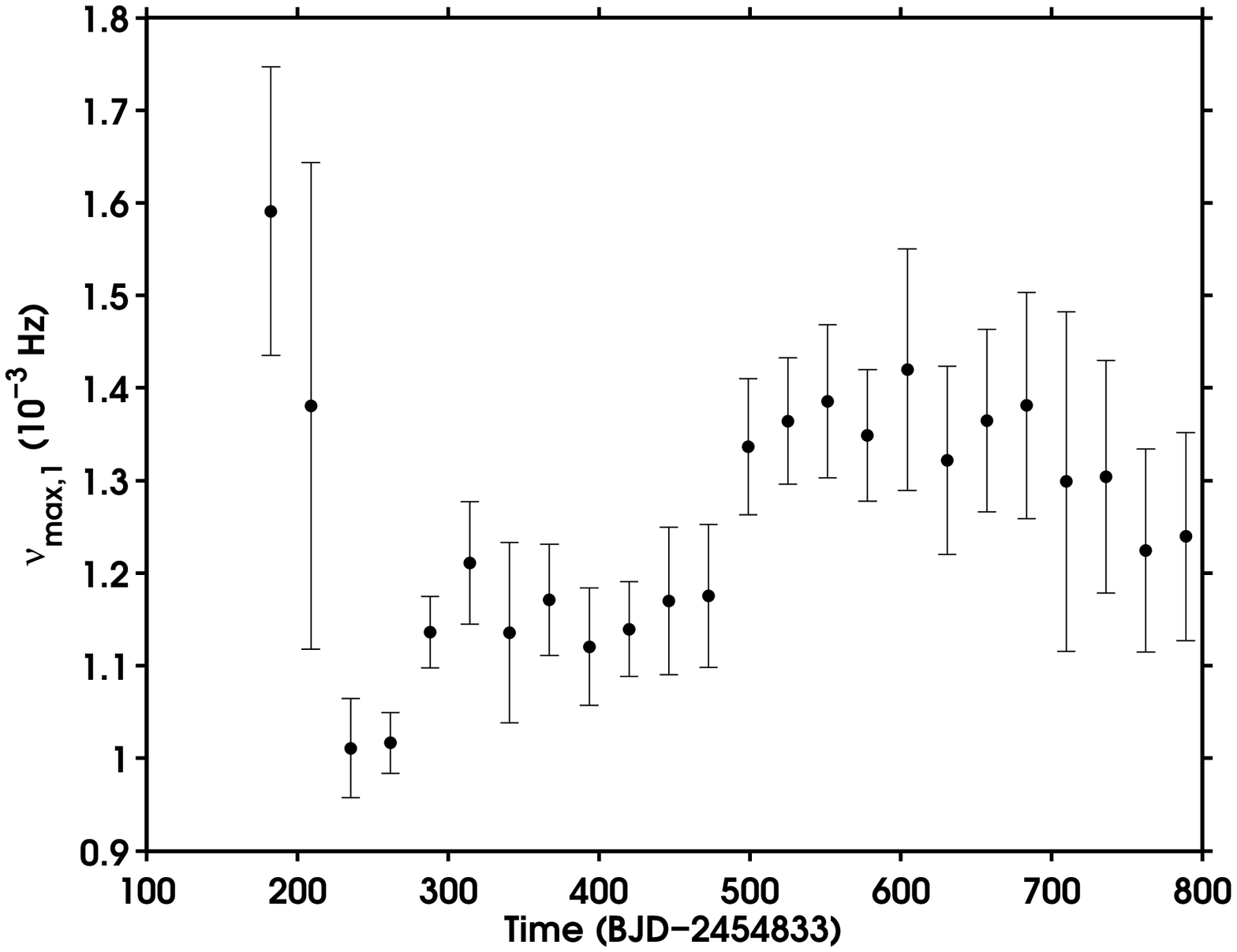} \includegraphics[width=0.48\textwidth]{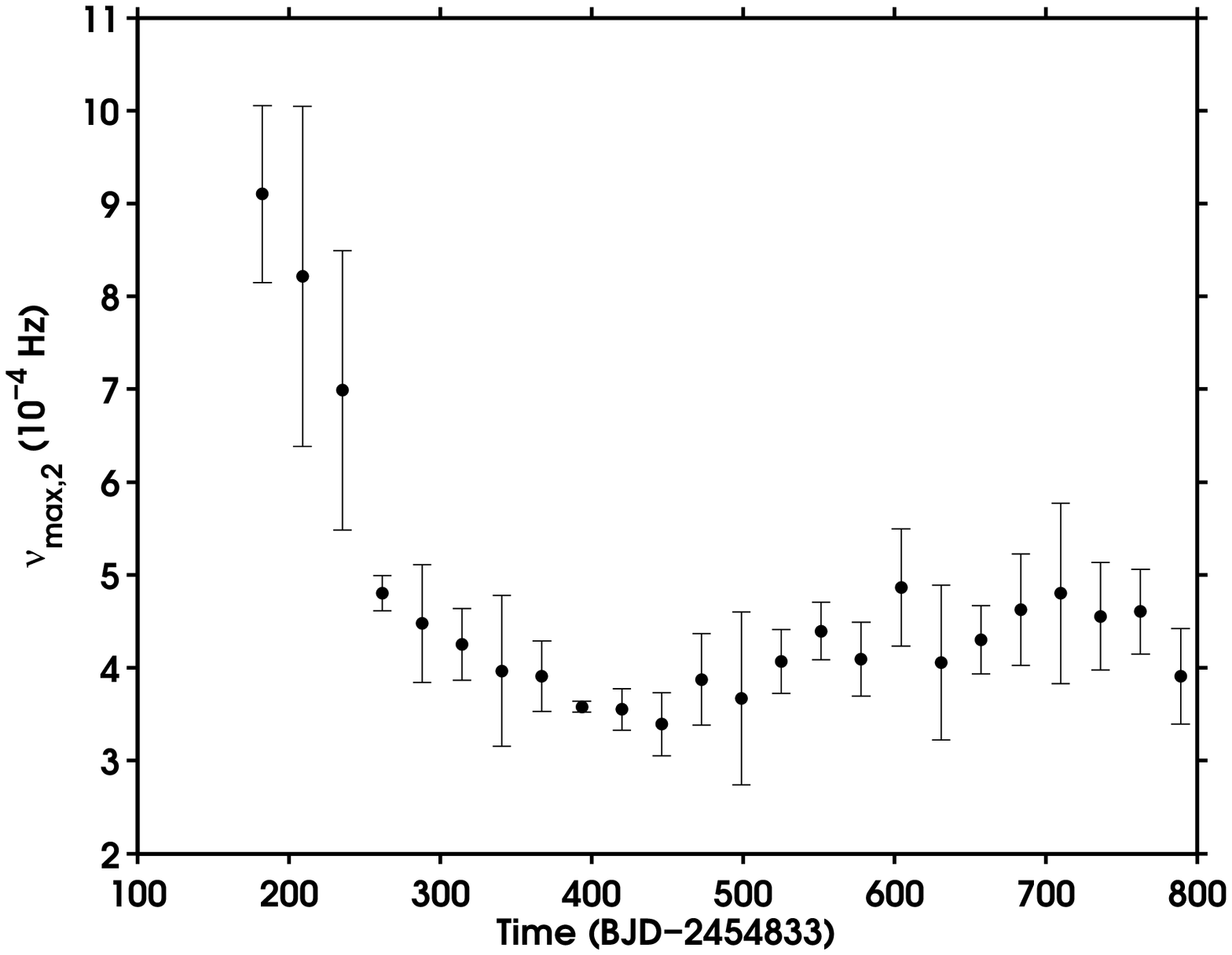}
\includegraphics[width=0.48\textwidth]{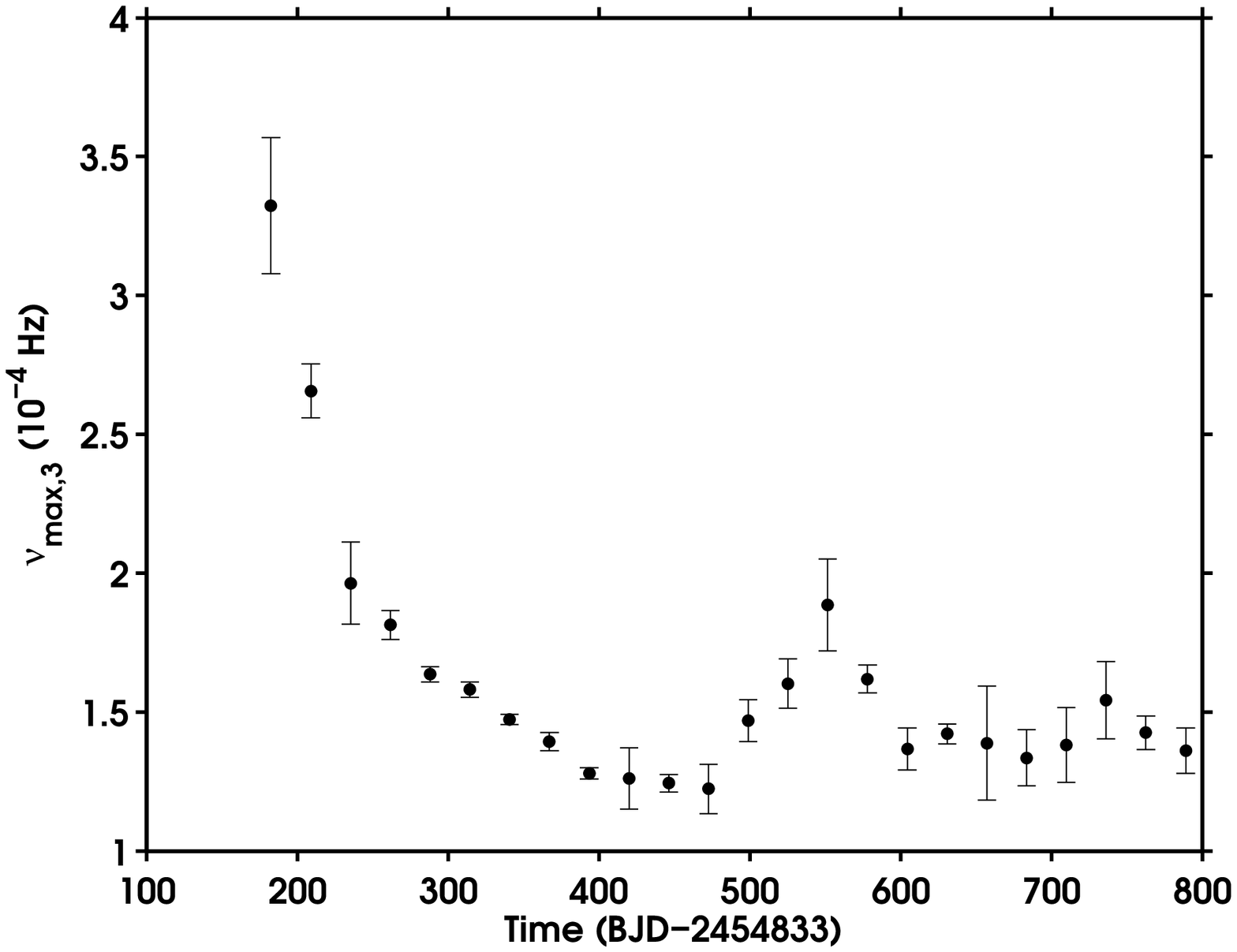} \includegraphics[width=0.48\textwidth]{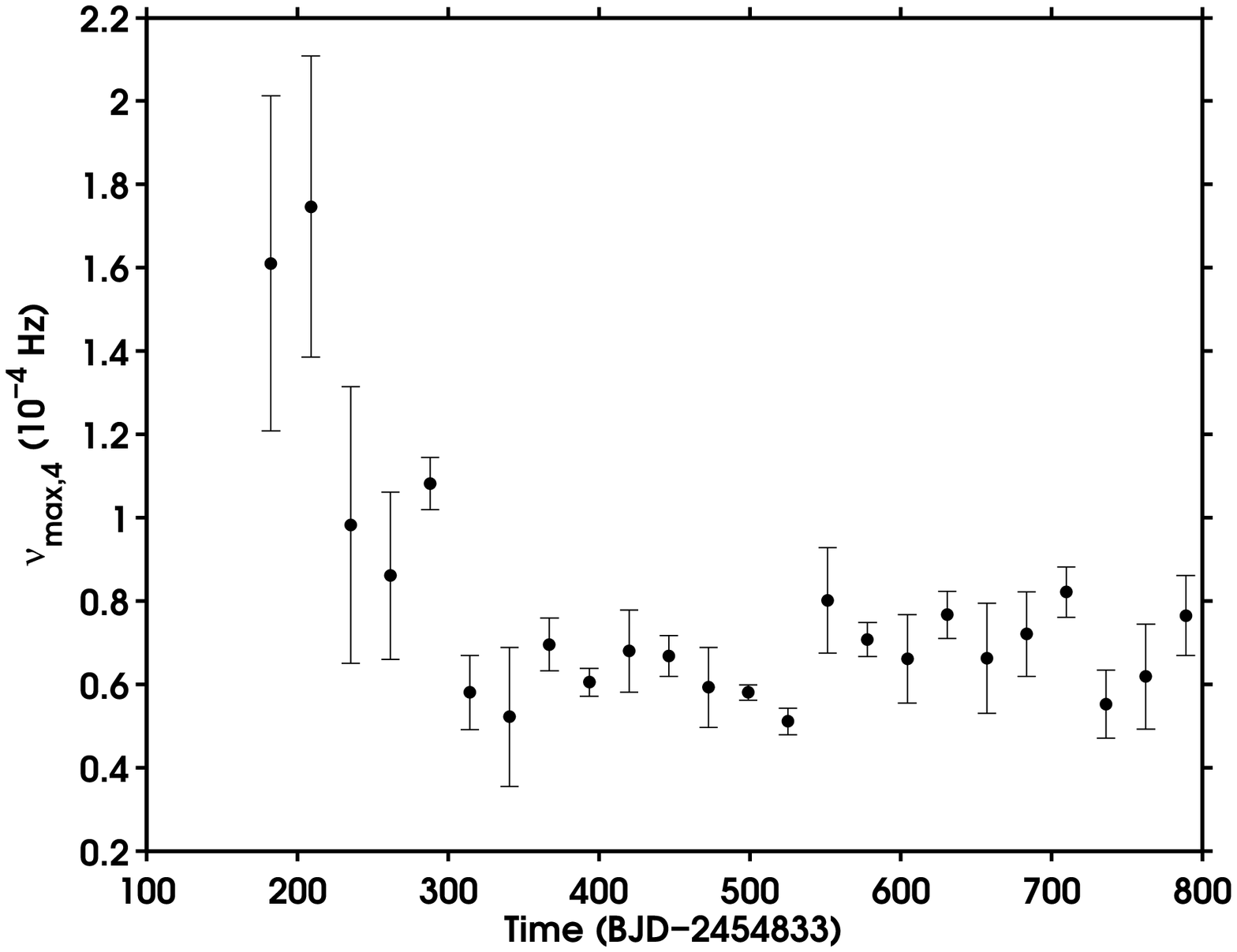}
\caption{Characteristic frequency, $\nu_{max}$, versus observation time for the four Lorentzians $L_1$, $L_2$, $L_3$ and $L_4$ from top-left to bottom-right respectively.}
\label{fig:5}
\end{figure*}

\begin{figure}
\includegraphics[width=0.5\textwidth]{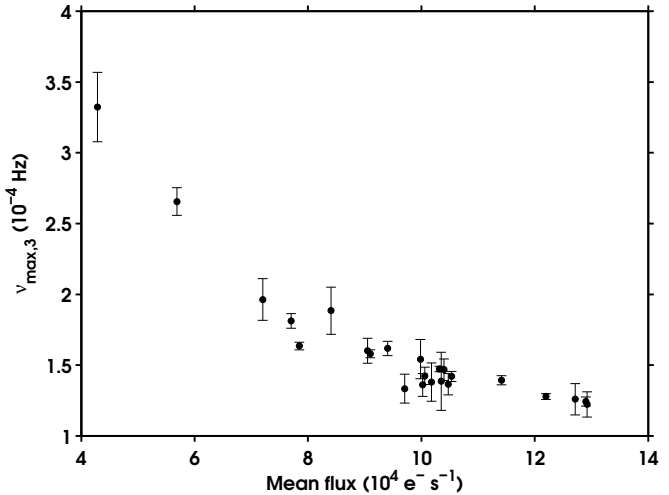} 
\caption{Correlation between the characteristic timescale $t_{max}$ of the Lorentzian $L_3$ and mean source flux.}
\label{fig:6}
\end{figure}

We searched for possible correlations within the fit parameters and both observation time and mean source flux. In this respect, interesting trends with observation time are found within the characteristic frequencies of the Lorentzians $\nu_{max}$, shown in Fig. \ref{fig:5}. The most prominent correlation however is found between the characteristic frequency of Lorentzian $L_3$ and the mean source flux, which obtained a Spearman correlation coefficient of $-0.89$, and is shown in Fig. \ref{fig:6}. To date, no similar correlation to that observed here has been reported for a CV, and we will discuss the possible origin of such variability in the next section. No other fit parameters display a clear trend with either flux or time.  

\begin{figure}
\includegraphics[width=0.5\textwidth]{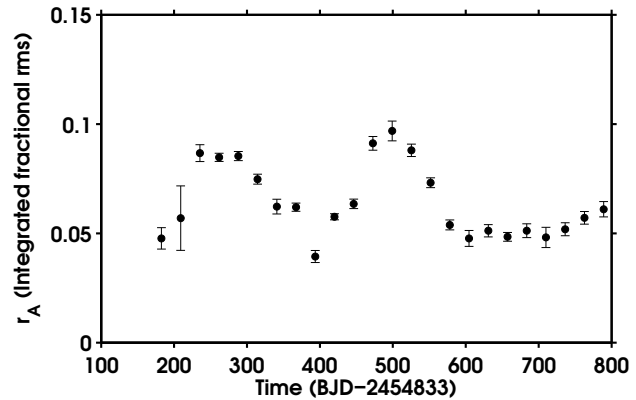} 
\caption{Evolution of the integrated fractional rms $r_1$ of the Lorentzian $L_1$ as a function of observing time.}
\label{fig:7}
\end{figure}

The high-frequency break, modelled by the Lorentzian $L_1$, is seen to slightly change in frequency, but does not correlate with mean flux (see top-left panel of Fig. \ref{fig:5}). However, one interesting trend within the Lorentzian $L_1$ arises by looking at the integrated fractional rms, $r_A$, as a function of time, shown in Fig. \ref{fig:7}. The two peaks at $\approx250$ and $\approx500$ days coincide with a small plateau and a decrease in flux within the MV Lyrae lightcurve respectively. Both these features, and the break frequency, will be discussed in more detail in the next section.

\begin{figure*}
\includegraphics[width=0.95\textwidth]{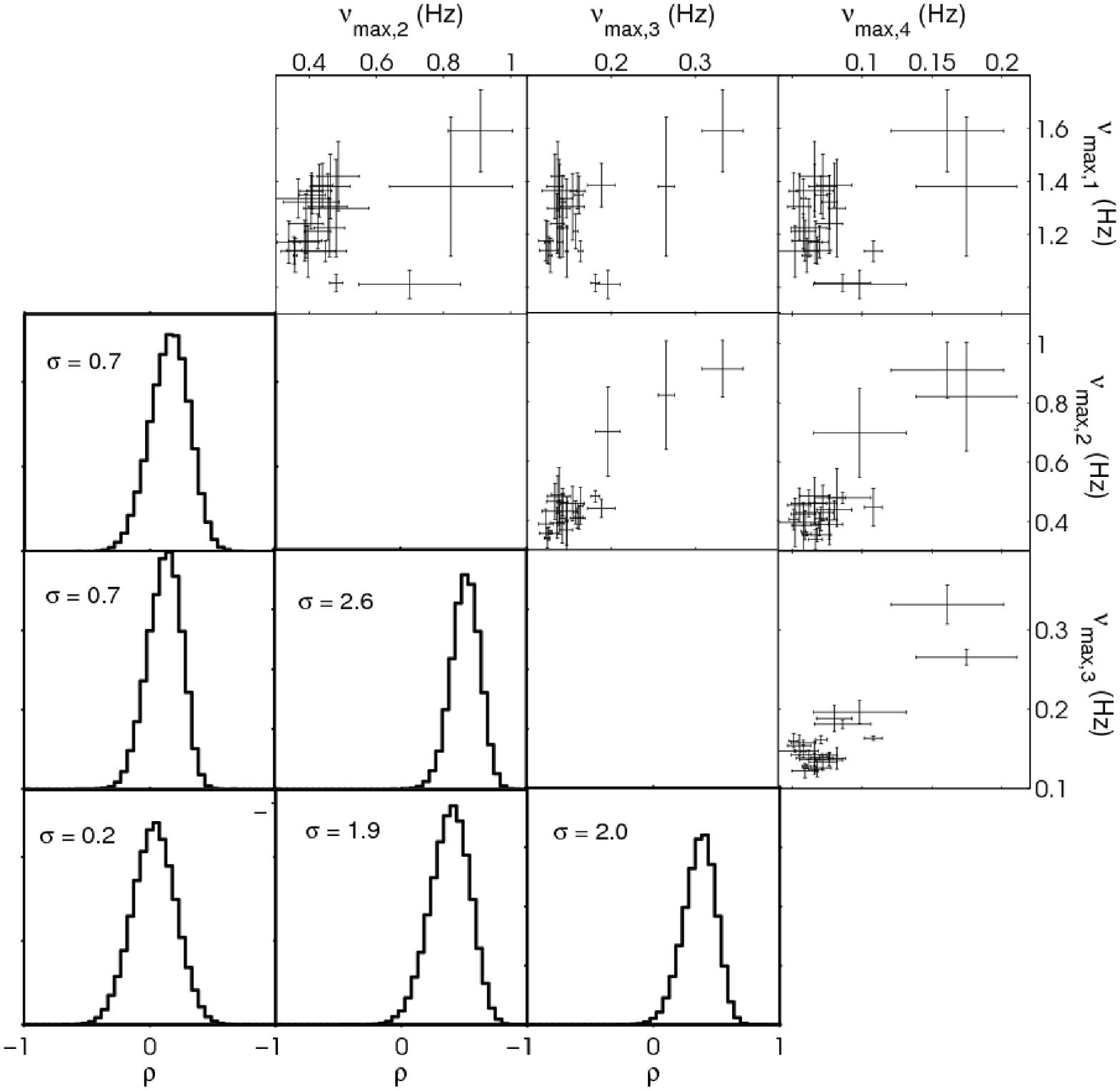} 
\caption{The top-right half of the figure shows the characteristic frequencies of the fitted Lorentzians. The bottom-left part of the figure shows the corresponding Spearman rank correlation coefficient values $\rho$ obtained from the analysis described in section \ref{sec:results} together with the corresponding correlation significance. Each panel matches its reflection along the diagonal.}
\label{fig:8}
\end{figure*}

As mentioned in the introduction, correlations between some of the Lorentzian components are observed in XRBs and some CVs. Specifically, the two-QPO correlation diagram produced for XRBs (\citealt{belloni}) and for CVs (\citealt{warner_qpo}) can be described by $P_{QPO}/P_{DNO}\approx15$. We have searched for a similar correlation between the 4 fitted Lorentzians in MV Lyrae. In Fig. \ref{fig:8} we show plots for the characteristic frequencies of all four fitted Lorentzians. In order to determine whether any of these pairs is statistically correlated we performed a Monte-Carlo correlation analysis similar to that used in \cite{scaringi_mCV}.

For each pair of characteristic frequencies we created two datasets. In the first (set A) we replaced each of the values with a random variable drawn from a normal distribution whose mean is equal to the fitted characteristic frequencies and whose standard deviation is equal to the respective errors. In the second (set B) we additionally shuffled the order of one of the two frequencies, thus randomising the correlation rank. We produced 200,000 samples for each set, and computed the correlation coefficient $\rho$ for each. The mean values of the $\rho$ distributions obtained from set A will then yield the \textit{intrinsic} correlation coefficients, taking into account the errors. The corresponding set B distribution will instead yield the width of the $\rho$ distribution in the absence of any correlation (which is centred at $0$). The statistical significance for each pair is then found by determining where the obtained mean $\rho$ values from set A fall on the obtained distribution of our mock dataset B. This analysis thus allows us to determine the correlation significance of the observed characteristic frequency pairs by taking into account the errors on each observation.

In Fig. \ref{fig:8} we show in the top-right panels plots for the characteristic frequency pairs, whilst in the bottom-left we show the resulting $\rho$ distribution from our first simulation and the corresponding correlation significance. None of the Lorentzian pairs are observed to display a clearly statistically significant correlation. However, it is still possible that some pairs may be correlated, and the low-frequency cluster of datapoints is biasing our observation. For example, the pair $\nu_{max,3}$ vs. $\nu_{max,2}$, which obtained the highest significance ($\sigma=2.6$) in our analysis, as well as some other pairs, do suggest this. However, further measurements are required at the highest frequencies in order to determine if this is the case. If the $\nu_{max,3}$ vs. $\nu_{max,2}$ pair were to be correlated after all, our analysis suggests they would follow the relation $t_{max,3}/t_{max,2}\approx3$, different from the known $P_{QPO}/P_{DNO}\approx15$ relation (\citealt{belloni,warner_qpo}).

\section{Discussion}\label{sec:discussion}

Here we discuss the implications of the observed QPO/broad-band noise component frequencies with regard to the standard, optically thick, geometrically thin accretion disk model (\citealt{ss_73}). Given the low binary inclination of $i=11^o-13^o$ (\citealt{SPT95}), the WD magnitude seen in the faint state (\citealt{hoard}) and the result from Paper~I, we are confident that most of the emitted light observed in the \Kepler\ passband is originating from the accretion disk. In what follows we will always assume $M_{WD}=0.722M_{\odot}$ (\citealt{hoard}), and a WD radius of $\approx7\times10^8$ cm.

The observed characteristic frequencies are usually associated with the dynamical or viscous timescale (\citealt{revn1,ID1,ID2,uttley1}). While it is not yet clear if one of these two timescales is the correct one, most other time-scales of interest (such as thermal) lie between these two. We will thus discuss them as likely limiting cases. The dynamical timescale is given by
\begin{equation}
\nu_{dyn}(r) = {1 \over t_{dyn}(r)} = \sqrt{{ GM \over r^{3}4\pi^2}} ,
\label{equ:5}
\end{equation}
\noindent and relates to the viscous timescale,
\begin{equation}
\nu_{visc}(r) = \alpha (H/R)^{2} \nu_{dyn}(r) ,
\label{equ:6}
\end{equation}
\noindent where $r$ is the accretion disk radius, $H/R$ and $\alpha$ are the disk scale height and viscosity parameter respectively, and $M$ is the mass of the central compact object (see e.g. \citealt{FKR}). 

For XRBs and AGN the observed broad-band noise variability is generally associated with the viscous timescales over a range of radii (\citealt{arevalo,ID1,ID2}). This is motivated by observations of the rms-flux relation seen in many XRBs and AGN (\citealt{uttley1}), which imply that the fast variability must be coupled to the slow variability by multiplicative processes (i.e. the resulting variability at the inner-most edges of the disk is the product of the variability produced at larger annuli). This can naturally arise as a result of the viscous propagation of material inwards through the disk. The standard model which seeks to reproduce these observations is the fluctuating accretion disk model (\citealt{kotov,lyub}), which can correctly reproduce the rms-flux relation as well as the general $1/f$ shape and high-frequency break observed in most PSDs of XRBs and AGN. In both cases the resulting inner-edge of the disk is inferred as being geometrically thick in order to allow the variability to propagate inwards without being damped. On the other hand, some authors have instead associated the break frequency observed in CVs to the dynamical timescale (Eq.\ref{equ:5}, \citealt{revn2,revn1}) of the inner-most edge of the disk. This association seems to alleviate the problem of a thick disk (since for this interpretation the disk can be either thin or thick), but instead suggests a large disk truncation radius at $\approx10R_{WD}$, and evaporation (\citealt{meyer}) is invoked as a mechanism to deplete the inner disk. Recently the rms-flux relation has also been observed MV Lyrae (\citealt{scaringi}) with the same data as that used in this work, suggesting that the mechanism responsible for mass transfer within the disks, and the associated variability, is similar for CVs, XRBs and AGN, whether this is viscous, dynamical or thermal. The main difference between the rms-flux relation observed in XRBs and that observed in the CV MV Lyrae are the timescales involved: for XRBs the rms-flux relation is observed in X-rays on timescales of seconds whilst for MV Lyrae it is observed in optical light on minute timescales. This timescale difference can potentially be attributed to the different emitting regions, since the variability observed in X-rays in XRBs is originating deeper within the potential well (thus faster variability) of the accreting compact source when compared to MV Lyrae. The study performed in Paper~I concentrated on analysing the high-frequency properties of MV Lyrae, whilst here we concentrate on the broad-band variability properties, and assess whether the observed features are phenomenologically similar to those in XRBs.

Given that the nature of variability is still debated, but recent phenomenological similarities seem to suggest a common origin for CVs, XRBs and AGN, we will, in what follows, associate the characteristic frequencies observed in MV Lyrae to either the viscous or dynamical timescales, and infer values for the accretion disk structure. This will allow us to compare our results to those inferred from X-ray studies of XRBs and AGN. In the next subsection we will specifically concentrate on the dynamical/viscous interpretations of two Lorentzians, namely $L_1$ and $L_3$. The general conclusions drawn from these two components will also hold for the other two Lorentzians as well, and we choose to discuss only $L_1$ and $L_3$ as these components seem to display the most interesting trends in Fig. \ref{fig:5}, \ref{fig:6} and \ref{fig:7}.

\subsection{Viscous interpretation}

\subsubsection{The low frequency Lorentzian $L_{3}$}\label{sec:vL3}

Here we discuss the possibility that the observed characteristic frequency $\nu_{max,3}$ is associated with the viscous timescale ($t_{visc}$) at some disk radius. The apparent anti-correlation with flux seen in Fig. \ref{fig:6} could then be suggesting a change in the viscosity parameter $\alpha$ and/or scale height $H/R$. Observationally, we find that the low frequency Lorentzian $L_{3}$ produces an rms variability of $\approx5\%$. As at least that fraction of the luminosity needs to be emitted in the variable emission region, we can first obtain a lower limit on the size of this emitting region. 

Measured temperature profiles of CVs with a similar accretion rate seem to follow the theoretical predictions of standard accretion discs up to $\approx 0.05R_{\odot}$ (\citealt{groot2}). Closer in, there is some controversy on the shape of the temperature profile, especially for the subclass of CVs SW Sex stars of which MV Lyrae is a member of. \cite{groot1,groot2}, have determined a flat-top profile through indirect eclipse mapping, however \cite{knigge00,knigge04} have later confirmed the possibility that in the high-luminosity state, at least some SW Sex stars posses a self-occulting component. This could be due to the presence of a thick disk, and would thus result in an observational bias when measuring the temperature profiles in the inner-most regions of the accretion disk, producing a flat-top temperature profile. 

\begin{figure}
\includegraphics[width=0.5\textwidth, height=0.40\textheight]{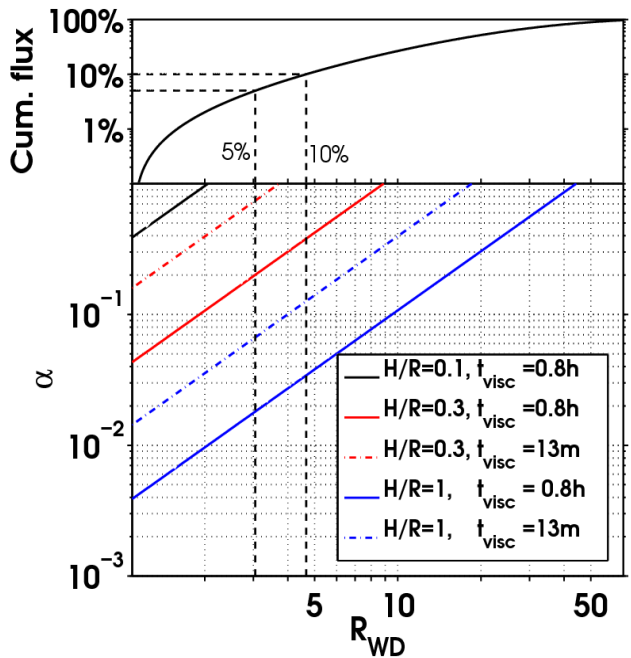}	
\caption{Top: Cumulative radial emission profile for the accretion disk in MV Lyrae folded through the \Kepler\ passband. Bottom: Inferred accretion disk radius and viscosity parameter $\alpha$ for different disk scale heights. The estimates have been obtained by associating the characteristic frequencies of 0.8 hours ($L_3$, solid lines) and 13 minutes ($L_1$, dashed lines) to the viscous timescale at a particular disk radius. The inferred $\alpha$ and $H/R$ values are found to be too large to be accommodated by a standard thin disk. The vertical black dashed lines indicate the minimal extent of the accretion disk given the observed $5\%$ and $10\%$ rms amplitude of the variability.}
\label{fig:9}
\end{figure}

Because of these controversies we will take the conservative limit and model the minimal size of the emitting region by using the standard disk temperature profile from \cite{FKR} (Eq.~5.49) with a mass accretion rate of $3\times10^{-9}M_{\odot}yr^{-1}$ (\citealt{linnell}). We then summed the emitted blackbody spectra produced at each disk annulus and folded this with the \Kepler\ response function\footnote{http://keplergo.arc.nasa.gov/CalibrationResponse.shtml}, producing the relative cumulative flux emitted by the disk as a function of radius in the top panel of Fig. \ref{fig:9}. The figure shows, for example, that if we observe say $\approx5\%$ rms variability (as observed for $L_3$), the emitting region must be larger than $\approx3R_{WD}$. In fact, $3R_{WD}$ in this scenario is a conservative lower limit for two reasons: (i) the absolute amplitude of the variability is larger than $\approx5\%$, and (ii) some CVs (namely SW Sex, \citealt{groot2}) show that the disk temperature profile flattens out at larger radii ($\approx0.2R_{\odot}$). Both these effects will push the minimal size of the emitting region to larger radii than those shown in Fig. \ref{fig:9}, and further increase the resulting limits on $\alpha$ and $H/R$.  

Having produced an estimate for the minimal size of the variable emitting region, we now place limits on the disk scale height ($H/R$) and viscosity ($\alpha$) for the fastest measured timescales of the Lorentzian $L_3$ of 0.8 hours. This is done as a function of radius in the bottom panel of Fig. \ref{fig:9} (solid lines). The inferred values for $\alpha$ and $H/R$ are too high to be accommodated by the standard thin disk model, where $\alpha\le0.1$ and $H/R\le10^{-2}$ are usually taken. Additionally, the most conservative values for $\alpha$ and $H/R$ (solid black line) are ruled out by the constraints imposed by the minimal size of the emitting region. Thus, if one wants to adopt the viscous interpretation, our result seems to suggest the presence of a geometrically thick accretion disk in order to explain the observed characteristic frequencies of the $L_3$ Lorentzian. This would be in agreement with the work of \cite{chur}, who have noted that for the fluctuating accretion disk model to work (\citealt{lyub,kotov}) and produce the observed rms-flux relation (\citealt{arevalo,scaringi}), one needs a geometrically thick disc.  

\subsubsection{The high frequency Lorentzian $L_{1}$}
We find the characteristic frequency of the Lorentzian $L_1$ to slightly vary throughout the observation, but the data does not allow us to determine whether it correlates with flux. In Paper~I (Fig.~7) we inferred values of $H/R\approx0.1$ for $\alpha\approx1$ assuming the $\approx13$ minute break originates from very close to the WD surface. Here we extend that analysis, and show in Fig. \ref{fig:9} (dashed lines) the dependence on both $H/R$ and $\alpha$ for various disk radii using the inferred PSD break at about 13 minutes to represent the viscous timescale. In this case the minimal size of the emitting region is determined by the $\approx10\%$ rms variability observed in Fig. \ref{fig:7}. Like the situation with the $0.8$ hour Lorentzian (see section \ref{sec:vL3}), this last constraint rules out the most conservative values for $\alpha$ and $H/R$, suggesting a geometrically thick disk. Furthermore we note that the constraints inferred from Fig. \ref{fig:9} are likely to be even larger than those inferred here. From the continuum variability modelling of accretion disks performed by \cite{ID1,ID2} we know that the true viscous timescale at the inner-most edge of the disk is even faster than that observed from the PSD break, pushing the inferred $\alpha$ and $H/R$ to even higher values.

One other important result concerning the high-frequency Lorentzian $L_1$ is the change in fractional rms variability shown in Fig. \ref{fig:7}. From the work of \cite{geertsema}, we can use the fractional rms as an indication of both $\alpha$ and $H/R$. This method is based on the effects of magneto-hydrodynamic (MHD) turbulence in an accretion disk, and associates the source of flickering to the energy dissipated per unit area at the disk surface. With this method \cite{baptista04} have inferred $\alpha\approx0.16$ for $H/R=10^{-2}$ in the CV system V2051 Oph, and comment on this result as being uncomfortably high to be accommodated by the standard thin disk model. In Paper~I we also showed how $\alpha$ varies between $\approx0.16$ and $\approx0.62$, and invoked an even thicker disk of $H/R\approx0.1$ to avoid at least the highest $\alpha$ values. The peaks of the rms variability presented in Fig. \ref{fig:7} coincide with a short-duration (few weeks) plateau observed at $\approx250$ days, and a drop in flux seen at $\approx500$, both shown in Fig. \ref{fig:1}). The behaviour observed in Fig. \ref{fig:7} could be suggesting a change in the geometry and/or viscosity parameter during the observation since the rms variability is tied to both $\alpha$ and $H/R$ in the work of \cite{geertsema}. If this were the case however, the inferred range for both $\alpha$ and $H/R$ is quite large ($0.02<\alpha<0.2$ and $0.01<H/R<0.1$) during the rms change between $5\%$ and $10\%$, and we caution interpretation of this result as it is not yet clear what physical mechanism is required to explain the observed phenomena.

\subsection{Dynamical interpretation}

\subsubsection{The low frequency Lorentzian $L_{3}$}
We can can alternatively try to associate the obtained characteristic frequencies of the Lorentzian $L_3$ from Fig. \ref{fig:5} with the Keplerian frequency at a specific disk radius. In order to explain the correlation seen in Fig. \ref{fig:6}, the emitting region would then radially grow in size, and increase in temperature, as the timescale changes from 0.8 to 2.1 hours.

Using the highest and lowest frequencies measured for the Lorentzian $L_3$, we can infer using Kepler's third law an accretion disk size ranging from $\approx0.39R_{\odot}$, for the highest frequency, up to $\approx0.74R_{\odot}$ for the lowest. These estimates, especially the low frequency one, are not consistent with the estimated position of the L1 point in this binary system, which sits at about $0.57R_{\odot}$ (\citealt{FKR}). One way to overcome this inconsistency would be to assume that the observed frequency is a beat between the Keplerian outer disk frequency ($\nu_{disk}$) and the orbital frequency ($\nu_{orb}$) of $3.19$ hours, $\nu_{beat}=\nu_{disk}+\nu_{orb}$. In this case, the outer disk-edge would be observed to grow from $\approx0.33R_{\odot}$ to $\approx0.53R_{\odot}$, consistent with the disk radially growing in size up to the L1 point. However the other Lorentzians, especially the lowest frequency $L_4$ Lorentzian, would then require a different mechanism to drive them, as their beat frequency would be too low to be accommodated within the L1 point.

\subsubsection{The high frequency Lorentzian $L_{1}$}
The high-frequency PSD breaks observed in accreting compact objects have also been associated with the Keplerian timescale at the inner-most edge of accretion disks. Specifically, \cite{revn1,revn2} have associated the PSD breaks observed in some magnetised WDs, NSs and the well-known CV SS Cyg to the dynamical timescale at the inner-most edge of the accretion disk. They inferred relatively large inner disk truncation radii for all systems they studied, and for SS Cyg in particular they inferred a truncation radius of $\approx10R_{WD}$. In order to deplete the inner disk \cite{revn2} suggest evaporation (\citealt{meyer}) as a mechanism to truncate the thin disk away from the WD surface. 

The break frequency observed in SS Cyg is very similar to the one observed here in MV Lyrae, and it is possible that we also observe the Keplerian frequency at the inner-most edge of the accretion disk truncated at about $\approx10R_{WD}$. We note, however, that dynamical processes will be more easily damped, as compared to viscous process, whilst they propagate through disk. This is because viscous damping will suppress the fast dynamical variability but not the slower viscous variability. As a result dynamical effects are observed as additive process in lightcurves, whilst viscous ones as multiplicative. The presence of the rms-flux relation at high frequencies thus suggests viscous interactions as the main source of variability, since additive processes would destroy the observed rms-flux relation. 

\section{Conclusion}
We have presented an analysis of the broad-band frequency behaviour of the accreting WD MV Lyrae based on data obtained by the \Kepler\ satellite. We have shown how the complex PSD can be decomposed with a number of Lorentzian-shaped functions. We further searched for possible correlations between the characteristic frequencies, and found the first frequency varying QPO in a CV, where frequency is inversely proportional to mean source flux. 

The characteristic frequencies associated with the fitted Lorentzians were used to explore the origin of variable emission in terms of viscous or dynamical processes as the two limiting cases. In the former case we infer extremely high values of both disk scale height $H/R\ge0.3$, and viscosity $\alpha\ge0.1$, suggesting the existence a geometrically-thick disk. This result is potentially in line with the work of \cite{knigge00,knigge04}, which have suggested the presence of a self-occulting accretion disk close to the WD. In the dynamical case we instead infer a large disk truncation radius of $\approx10R_{WD}$, but the presence of other components undermines a dynamical interpretation, at least for the lowest frequency components. More importantly, the presence of the rms-flux relation observed in MV Lyrae (\citealt{scaringi}) and other XRBs/AGN (\citealt{uttley1}), potentially rule out dynamical effects as the source of variability (since the fast dynamical variability will be damped), but favours a viscous origin for the observed broad-band noise components, at least at the highest observed frequencies. 

In summary both viscous and dynamical (and consequently thermal) timescales struggle to consistently explain the observed broad-band variability in MV Lyrae. However, our analysis seems to suggest, at least phenomenologically, that the mechanisms which give rise to the observed rms-flux relation(s) and characteristic frequencies observed here and in other XRBs/AGN need to occur in all accretion discs (whether they are thin or thick), possibly suggesting a similar physical origin for the variability in both types of systems. 

The broad-band variability properties of MV Lyrae remain yet to be fully understood. More generally, the variability properties of both CVs and XRBs also remain an enigmatic observational feature. Although both types of systems possess an accretion disk which is in many ways similar, there has not yet been enough observational data to provide a complete comparative study of the broad-band variability properties. However, it is clear that the phenomenological properties between the broad-band variability observed in MV Lyrae and in X-ray binaries and Active Galactic Nuclei are very similar (when appropriately scaled), and are potentially driven by a common accretion mechanism.

\section*{Acknowledgements}
This paper includes data collected by the \Kepler\ mission. Funding for the \Kepler\ mission is provided by the NASA Science Mission directorate. This research has made use of NASA's Astrophysics Data System Bibliographic Services. S.S. acknowledges funding from NWO project 600.065.140.08N306 to P.J. Groot. M.S. acknowledges funding from the NASA grant NNX11AB86G. S.S. wishes to acknowledge G. Nelemans and A. Achterberg for useful and insightful discussions.

\bibliographystyle{mn2e}
\bibliography{MVLyr_paper}

\label{lastpage}

\end{document}